# Nonmonotonic Relaxation as a Result of Spatial Heterogeneity in the Model of In-series Blocks Chain

A.A. Bedulina, A.V. Kobelev

*Institute of physics of metals UrD Russian Academy of Sciences, Yekaterinburg, Kovalevskoy 18*

Recently the materials possessing structure of molecular and supramolecular matrix are more and more actively studied. They are relative to many polymeric materials of a technological origin, such as rubber, and living biological tissues [1]. Processes of mechanical deformation of such continuous media have peculiarities connected, first, with accounting for internal friction and dissipation of energy, and secondly, with nonlinearity of their elastic and viscous properties, that is with violation of Hook' and Newton's laws. Problem of modeling of these systems reduce to the analysis of the corresponding equivalent mechanical or electric circuit (see examples in the classical monograph [2]). Rheological properties [3-4] of described medium are governed by the differential equations of the first order on time (the evolution equations), as well as a huge variety of other physical processes. The physical phenomena in nonlinear systems with dissipation have a big community, including such it would seem far areas, as dynamics of magnetization in ferrite [5]. Therefore the problem of studying new effects of viscous friction in the conditions of nonlinearity and heterogeneity, is very actual as in respect of fundamental research nonlinear and non-uniform environments, and in many areas of materials science, design of new materials, engineering of biological substitutes of living tissues and development of the micromagnetic devices using essentially new opportunities. Traditional approaches to mechanics of viscoelastic bodies sometimes are excessively difficult [6], and more evident and available representations are necessary. The invaluable role in studying of the operating processes mechanisms of elastic deformation and motility of biological materials, is played by the mathematical modeling [7].

## Introduction

In the present work the phenomenological approach is applied to a problem of modeling viscoelastic properties of structured materials based on representation of tissues' elementary blocks by two-dimensional (and three-dimensional) network of linear elastic and viscous components. This approach have been developed in common group of researchers of Institute of immunology and physiology and Institute of metal physics of UrD Russian Academy of Sciences within the program of nonlinear mechanics research [14, 19]. As a matter of fact, the model is represented mathematically by the system of differential equations of the first order on time with the number of derivatives coinciding with the number of viscous Newtonian elements in the model. Nonlinearity of viscoelastic properties is provided with the effects of change the direction of inclined elements in relation to a longitudinal axis at longitudinal deformation. Masses of elements and friction in knots of their connection are considered negligible therefore are excluded from consideration. At an appropriate choice of the values of elastic and viscous parameters of elements it is possible to reproduce such nonlinear effects known from experiment, as change in slope of a stationary "stress - deformation" curve (effect of "toughening"), the dependence of a relaxation time on the current level of deformation step, a "banana-like" form of a hysteresis loop, etc.

We use alike approach to research the influence of material heterogeneity on its mechanical properties. Heterogeneity is imitated by consecutive connection in a chain of blocks of a certain type (of "rhombic" model), possessing various values of the corresponding linear elements elastic and viscous parameters. Change of viscous and elastic parameters along a chain expresses heterogeneity of properties of a material as a whole.

As similar sets of the equations of the first order describe the most various objects from the point of view of their evolution, results of research are applicable in very many areas of nonlinear dynamics and possess sufficient degree of universality for their application in practice of materials science and designingof new materials.

**Historical review**

The main point of theoretical approach to research the mechanical properties of the non-uniform structured materials and, in particular, living biological tissues, is the necessity to combine characteristics of a solid body and liquid, for example, elasticity and fluidity in one model. Such task arose far before the basic provisions of viscoelasticity theory were formulated. Pioneer achievements in this area are Kelvin-Voigt and Maxwell's models which represent, respectively, parallel and consecutive connection of elastic and viscous linear elements, that is the Hookean spring and the Newtonian damper [2, 4, 8]. Many extended protein molecules which are imitated by springs when modeling, really have straight-line Hookean characteristics in the certain range of deformations [9] whereas other proteins incorporating globular regions (for example, titin [10-11]), have nonlinear, close to exponential, the dependence "force length" explained with their statistical nature [12]. Let's note that Kelvin's model representing, as a matter of fact, a spring, shipped in the viscous liquid providing losses on friction at its deformation, most naturally considers such association of elastic and viscous properties. However in this model it isn't possible to describe stress relaxation at constant deformation since it isn't suitable for the description of sharp jumps of deformation. The Maxwell's model gives such opportunity and is suitable for the description a viscous current ("creep" [13]) of viscoelastic materials. Transition from these two-component models to models with additional elastic or viscous elements [8, 14] allows expanding considerably the range of the modeled phenomena and materials. The standard one-dimensional model possessing essential universality is the Frenkel's model and its versions [2, 15].

As all one-dimensional models of any complexity combined of linear elements, have linear response function ("force - length" and "force - speed"), it raises a problem of the description of nonlinearity effects, which are immanent to considered tissues and are frequently observed experimentally. For the first time in [16] the two-dimensional combinations of the linear elastic elements are suggested, allowing nonlinearity of a curve "force - length" in the model at the expense of new degree of freedom which is defined by transversal deformations. Conditions for values of parameters in various on topology (rhombic) models providing effect of "toughening" of a material at its deformation are found. The effect arise, because in the beginning less rigid transverse elastic elements work, and then more rigid inclined do. In [17] viscosity was considered by shunting by dampers of some elastic elements of rhombic model within Kelvin-Voigt approach. Thus it was possible to reproduce characteristic of many biological soft tissues a "banana-like" form of a hysteresis "force - length" curve at slow periodic deformation. By a form of this dependence it is suggested to estimate which of a model elements possess large viscosity that would allow drawing out certain conclusions on parameters of internal structure of examined tissue samples. It is represented interest also by results of calculations of a relaxation of deformation in conditions when during action of periodic loading the relaxation of deformation of less viscous elements comes to the end whereas for more viscous proceeds. The relaxation time proportional to viscosity coefficient thus has to be greater than the deformation period.

The conclusion is important that in two-dimensional rhombic models with Kelvin-Voigt's blocks the force relaxation at constant deformation is described so simply and naturally, as well as in Maxwell's model, at the expense of transversal degree of freedom [18-19]. And it appears that the form of a force relaxation curve at spasmodic deformation (a step, a ladder, a rectangular impulse) is very close in a form to the Maxwell's model corresponding curve irrespective of the fact which the elastic element (inclined or transversal) possesses viscosity. It is obvious that in this case it is impossible to use models in which all elastic elements possess viscosity, owing to "frozen up" of such models at step deformation. The form of a force relaxation curve in reply to sharp step jump of deformation possesses big informational content in respect of a choice of adequate model and its verification on available experimental data [20]. Classical exponential nature of this dependence in Maxwell's model, in more complex multielement models with two and more viscosity parameters, than changes on two-exponential form usually observed on practical experience with quickly and slowly falling down components [14]. The detailed analysis of a force-relaxation-curve form is carried out in one-dimensional [20] and two-dimensional [18] models. Thus, the exit out from a framework of the one-dimensional models which are widely used for the analysis of a rheology of materials [21], opens ample opportunities of studying and application of materials with nonlinear characteristics though sometimes such models are used in obviously linear area [22].

It is interesting to note that a stress relaxation curve really observed in experiment on a papillary muscle of laboratory animals is close to exponential dependence with a power proportional to a time root [11]. Such special dependence, usually, takes place in materials with fractal structure, and there is the extensive area of research assuming a fractional type of derivatives on time in the equations of evolution in the elementary models of Kelvin and Maxwell [6, 23]. It is natural that at the expense of the fractional derivatives it is possible to receive a number of consequences for the response functions, explaining their nonlinear character. We won't concern here this interesting area because of a lack of information on a role of biological tissues fractal structure in their behavior.

The most informative in respect of identification and an explanation of mechanisms of nonlinearity is the dependence of stress relaxation time on current deformation level, i.e. on step number, with a step growth or falling off deformation course. Usually in experiment the relaxation time, i.e. duration of stress falling off, in process of the deformation growth (in rising step number) [10-11], increase is observed. It is explained by a more complete involvement of the rigid elastic elements possessing large viscosity, in process on the deformation growth. Similar nonlinear dependence of relaxation time is modeled quite naturally by means of idea of the inclined elements unfolding in a two-dimensional elastic model at longitudinal stretching, in the same way as well as effect of "toughening" of a stationary curve "stress - deformation".

Usually as a direct consequence of viscous friction and dissipation in a material the hysteresis phenomena at periodic loading are considered. Studying of a viscoelastic hysteresis "force - length" loop from a cycle to a cycle in various two-dimensional and three-dimensional rheological models allows to investigate the nonlinear effects [14, 17, 24] connected with a form and width of a hysteresis loop, and effects of evolution [25], i.e. changes of a curve position and shape.

It is necessary to emphasize especially a role of heterogeneity of studied samples. It is to a lesser extent characteristic for the artificial structured materials, and it is more essential to biological soft tissues where spatial heterogeneity of mechanical and viscous characteristics plays sometimes a key role in their functional stability and working capacity. As the first step in modeling of heterogeneity we use consecutive connection of blocks of models with different mechanical parameters, imitating thereby change of some properties (for example, stiffness) with longitudinal coordinate. The results received on model of ten consistently connected one-dimensional blocks of standard model of a myocardium (the Hill model) with the generator of force [26] are interesting. Nonlinearity of model is caused by complex character of direct and backward feedback mechano-electric interactions at excitement impulse generation. It appeared that in the described model there is an essential non-uniform parameters self-organization at the given delay time of consecutive elements periodic excitement.

Let's note that in the majority of the considered mechanical models of tissues the nonlinearity is weak and doesn't result in characteristic features for nonlinear dynamics like solitons, instabilities and bifurcations. In the language of phase portraits such systems are characterized by simple cycles of evolution. At the same time, in significantly nonlinear interacting non-uniform systems such unusual phenomena, as a nonmonotonic relaxation (for example, in [5]) can be observed. It means that in the initial stage in one part of nonlinear interacting system there takes place not a recession, and increase of some dynamic characteristics, for example, stress or deformations, and then its further full relaxation comes on. This area of the nonlinear phenomena is connected closely with the phenomenon of relaxation oscillations [14]. In this work the models of consequently connected rhombic blocks with various elastic and viscous parameters are investigated. The special attention is paid to the search of effects like a nonmonotone relaxation which besides basic fundamental significance can lead to their possible use for development of materials with unusual self-regulating viscoelastic properties.

**Description of the model base**

Modeling of viscoelastic properties of soft tissues is based on idea of a modular character of a biological tissue structure that is on existence of the elementary unit reflecting both morphological and functional properties of a material. Distinctive feature of biological tissues is nonlinear dependence of force on deformation by stretching with effect of "toughening" at sufficient deformations. At a choice of the model corresponding to structural unit of tissue, this feature can be considered not by the abstract assumption of material parameters (stiffness and viscosity) dependence on deformation, and by means of peculiarities of spatial structure of the model reflecting its morphology, following the approach suggested in [14]. Thus nonlinear dependence of force on deformation can be reproduced in models with linear (the Hookean and the Newtonian) elements and to have completely "geometrical" origin.

**1. Basic equations**

*1.1 . Rhombic model with inclined viscous elements*

As structural unit of a tissue the rhombic model with transverse elastic and inclined viscoelastic elements (fig. 1) was investigated. The inclined element represents Kelvin-Voigt rheological model in which the elastic element (spring) is connected in parallel to a damper reflecting viscous properties. The spring is characterized by length $l_1$ and stiffness (rigidity) parameter $K_1$, and a damper – length $l_2$ and viscosity parameter $H_1$. We use the assumption that lengths of an elastic and viscous element are equal $l_1 = l_2$, and we will present Kelvin-Voigt model in more evident look, as the elastic element placed in the liquid viscous tank. Force developed in model of Kelvin-Voigt at longitudinal deformation of an inclined element, is equal to the sum of elastic and viscous forces:

$$\vec{F_1} = K_1 \cdot (\vec{l_1} - \vec{l_{10}}) + H_1 \cdot \frac{d}{dt}\vec{l_1} \qquad (1.1)$$

$l_{10}$ - initial (reference) length of inclined element.

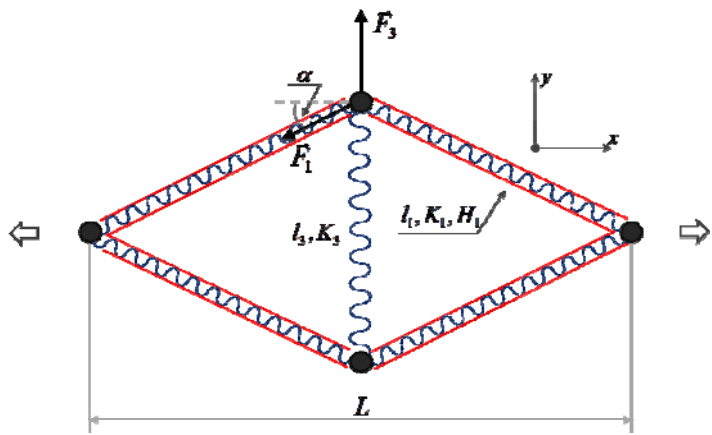

Fig. 1. Rhombic model with inclined viscous elements.

The first term represents Hook's law for a spring, the second – the Newton's viscosity law (internal friction) for a damper. Force developed by a transverse element of rhombic model:

$$\overline{F_3} = K_3 \cdot (\overline{l_3} - \overline{l_{30}}) \tag{1.2}$$

$l_3$ – length of a transverse element (spring), $l_{30}$ - the initial (reference) length of an element, $K_3$ – stiffness (rigidity) of a transversal spring.

Connections of elements of model we consider as hinges without friction, the mass of elements we don't consider. Balance of forces in knot of model describes the equation (fig. 1):

$$\overline{F_1} + \overline{F_3} = 0 \tag{1.3}$$

Let's put the equation in a projection to the OY axis:

$$F_3 - 2 \cdot F_1 \cdot \sin \alpha = 0 \tag{1.4}$$

$$\sin \alpha = \frac{l_3}{2 \cdot l_1}$$

$L$ - total longitudinal length of rhombic model (coincides with rhomb length).

Relation between the geometrical sizes of structural elements and total longitudinal length of the model reflects the equation:

$$L^2 + l_3^2 = 4 \cdot l_1^2 \tag{1.5}$$

Combining the received equations, it is possible to write down the system describing a statics and dynamics of rhombic model at longitudinal deformation:

$$\begin{cases} F_3 - 2 \cdot F_1 \cdot \sin \alpha = 0 \\ L^2 + l_3^2 = 4 \cdot l_1^2 \end{cases} \tag{1.6}$$

Expressions for modules of forces:

$$F_1 = K_1 \cdot (l_1 - l_{10}) + H_1 \cdot \frac{d}{dt} l_1 \tag{1.7}$$

$$F_3 = K_3 \cdot (l_{30} - l_3) \tag{1.8}$$

Substituting values of modules of forces and a sine of the angle between an inclined element and a horizontal line and carrying out transformations, we will receive system of the ordinary nonlinear differential equations for variable lengths of elements of model $l_1(t)$ and $l_3(t)$:

$$\begin{cases} K_3 \cdot (l_{30} - l_3) \cdot l_1 - (K_1 \cdot (l_1 - l_{10}) + H_1 \cdot \frac{d}{dt} l_1) \cdot l_3 = 0 \\ L^2 + l_3^2 - 4 \cdot l_1^2 = 0 \end{cases} \tag{1.9}$$

*2. Consecutive connection of two rhombic blocks with inclined viscous elements*

The following step comes when tissue modeling is a connection of structural units. Let's consider consecutive connection of two rhombic models with inclined viscous elements (fig. 2).

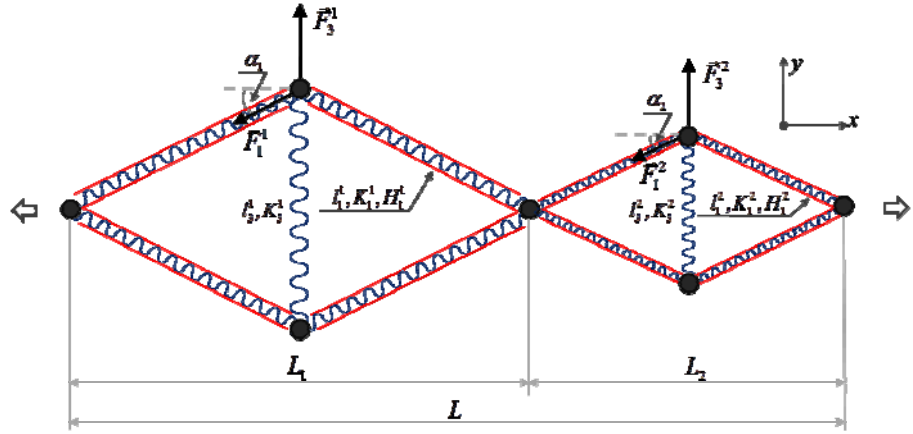

Fig. 2. Consecutive connection of two rhombic blocks with inclined viscous elements (doublet, tandem).

Let's consider the system of the equations characterizing consecutive connection of two rhombs. It is obvious that the subsystems describing dynamics of each of rhombs separately and including the equation of equality of forces in knot will enter it, connecting two Kelvin-Voigt blocks and a transverse elastic element, and the equation describing geometry of model in any time point.

The subsystem describing dynamics of the first rhomb in connection:

$$\begin{cases} F_3^1 - 2 \cdot F_1^1 \cdot \sin \alpha_1 = 0 \\ (L_1)^2 + (l_3^1)^2 = 4 \cdot (l_1^1)^2 \end{cases} \quad (1.10)$$

$$F_1^1 = K_1^1 \cdot (l_1^1 - l_{10}^1) + H_1 \cdot \frac{d}{dt} l_1^1$$

$$F_3^1 = K_3^1 \cdot (l_{30}^1 - l_3^1)$$

$$\sin \alpha_1 = \frac{l_3^1}{2 \cdot l_1^1}$$

where, $K_1^1$, $l_1^1$, $l_{10}^1$, $H_1$ – respectively rigidity, length, initial length and viscosity of an inclined element of the first rhomb representing one-dimensional model of Kelvin-Voigt; $K_3^1$, $l_3^1$, $l_{30}^1$ – respectively stiffness, length and initial length of a transversal elastic element of the first rhomb in consecutive connection; $L_1$ – rhomb length.

The subsystem describing dynamics of the second rhomb in connection has a similar appearance:

$$\begin{cases} F_3^2 - 2 \cdot F_1^2 \cdot \sin \alpha_2 = 0 \\ (L_2)^2 + (l_3^2)^2 = 4 \cdot (l_1^2)^2 \end{cases} \quad (1.11)$$

$$F_1^2 = K_1^2 \cdot (l_1^2 - l_{10}^2) + H_2 \cdot \frac{d}{dt} l_1^2$$

$$F_3^2 = K_3^2 \cdot (l_{30}^2 - l_3^2)$$

$$\sin \alpha_2 = \frac{l_3^2}{2 \cdot l_1^2}$$

where, $K_1^2$, $l_1^2$, $l_{10}^2$, $H_2$ – respectively rigidity, length, initial length and viscosity of an inclined element of the second rhomb; $K_3^2$, $l_3^2$, $l_{30}^2$ - respectively rigidity, length and initial length of a cross elastic element of the second rhomb in consecutive connection; $L_2$ – rhomb length.

Let's analyze the number of variables necessary for the description of dynamics of consecutive connection. If we consider rigidity $K_1^1$, $K_1^2$, $K_3^1$, $K_3^2$, viscosity $H_1$, $H_2$, initial lengths $l_{10}^1$, $l_{10}^2$, $l_{30}^1$, $l_{30}^2$ of elements, and the full length $L$ of consecutive connection as known quantities the variables are, $l_1^1(t)$, $l_1^2(t)$, $l_3^1(t)$, $l_3^2(t)$, $L_1(t)$ and $L_2(t)$. The number of variables is six, and so the system describing consecutive connection of rhombic models, has to contain not less than six equations.

One of the equations, supplementing couple of subsystems of each rhomb, describes the total length of consecutive connection $L$ as the sum of lengths of models of rhombs:

$$L_1 + L_2 = L \tag{1.12}$$

The last equation of system has to describe interaction of rhombs. For finding of the equation it is necessary to write down a projection of forces in the connection knot of rhombic models on the OX axis:

$$F_1^1 \cdot \cos \alpha_1 = F_1^2 \cdot \cos \alpha_2 \tag{1.13}$$

$$\cos \alpha_1 = \frac{L_1}{2 \cdot l_1^1}$$

$$\cos \alpha_2 = \frac{L_2}{2 \cdot l_1^2}$$

The equation describes interaction of rhombs, however, besides variables $l_1^1(t)$, $l_1^2(t)$, $L_1(t)$ and $L_2(t)$, it contains derivatives of lengths of inclined elements of models $\frac{dl_1^1}{dt}$ and $\frac{dl_1^2}{dt}$ therefore for the calculations it is more convenient to transform it. Let's express $F_1^1$ and $F_1^2$ from the first equations of subsystems through forces developed by cross springs $F_3^1$ and $F_3^2$:

$$F_1^1 = \frac{F_3^1}{2 \cdot \sin \alpha_1} = F_3^1 \cdot \frac{l_1^1}{l_3^1} \tag{1.14}$$

$$F_1^2 = \frac{F_3^2}{2 \cdot \sin \alpha_2} = F_3^2 \cdot \frac{l_1^2}{l_3^2}$$

Substituting the expression for forces developed by inclined elements in equation, and the expressions, connecting cosines of angles with lengths of elements of model, we will receive the equation:

$$F_3^1 \cdot \frac{L_1}{l_3^1} - F_3^2 \cdot \frac{L_2}{l_3^2} = 0 \tag{1.15}$$

Let's unite the specified equations in the following closed system:

$$\begin{cases} F_3^1 - 2 \cdot F_1^1 \cdot \sin \alpha_1 = 0 \\ (L_1)^2 + (l_3^1)^2 = 4 \cdot (l_1^1)^2 \\ F_3^2 - 2 \cdot F_1^2 \cdot \sin \alpha_2 = 0 \\ (L_2)^2 + (l_3^2)^2 = 4 \cdot (l_1^2)^2 \\ L_1 + L_2 = L \\ F_3^1 \cdot \frac{L_1}{l_3^1} - F_3^2 \cdot \frac{L_2}{l_3^2} = 0 \end{cases} \tag{1.16}$$

Having realized expressions for $F_1^1$, $F_1^2$, $F_3^1$, $F_3^2$, $\sin \alpha_1$, $\sin \alpha_2$, we will receive the system of ordinary nonlinear differential equations for variables $l_1^1(t)$, $l_1^2(t)$, $l_3^1(t)$, $l_3^2(t)$, $L_1(t)$ and $L_2(t)$, describing dynamics of consecutive connection of two rhombic models, i.e. model of a consecutive doublet (tandem):

$$\begin{cases} K_3^1 \cdot (l_{30}^1 - l_3^1) \cdot l_1^1 - (K_1^1 \cdot (l_1^1 - l_{10}^1) + H_1 \cdot \frac{d}{dt} l_1^1) \cdot l_3^1 = 0 \\ (L_1)^2 + (l_3^1)^2 - 4 \cdot (l_1^1)^2 = 0 \\ K_3^2 \cdot (l_{30}^2 - l_3^2) \cdot l_1^2 - (K_1^2 \cdot (l_1^2 - l_{10}^2) + H_2 \cdot \frac{d}{dt} l_1^2) \cdot l_3^2 = 0 \\ (L_2)^2 + (l_3^2)^2 - 4 \cdot (l_1^2)^2 = 0 \\ L_1 + L_2 - L = 0 \\ K_3^1 \cdot (l_{30}^1 - l_3^1) \cdot \frac{L_1}{l_3^1} - K_3^2 \cdot (l_{30}^2 - l_3^2) \cdot \frac{L_2}{l_3^2} = 0 \end{cases} \quad (1.17)$$

## 3. Consecutive connection of three rhombic blocks with inclined viscous elements

For the description of consecutive connection of three rhombic blocks (triplet) (fig. 3) with known parameters of rigidity, viscosity, initial lengths of elements and full length of model, it is necessary to know values of the following variables: $l_1^1(t)$, $l_1^2(t)$, $l_1^3(t)$, $l_3^1(t)$, $l_3^2(t)$, $l_3^3(t)$, $L_1(t)$, $L_2(t)$ and $L_3(t)$. The number of variables is nine, so the system of the equations describing consecutive connection of three rhombic models with inclined viscous elements, has to contain not less than nine equations. The equations of system can be collected by the following groups: three couples equations describing each rhomb separately (the equation of balance of forces in knot and the equation of geometry of a rhomb), the equation for the full length of model through the sum of lengths of each of rhomb $L_1$, $L_2$, $L_3$, and two equations describing interaction of rhombs (the first and the second, the second and the third).

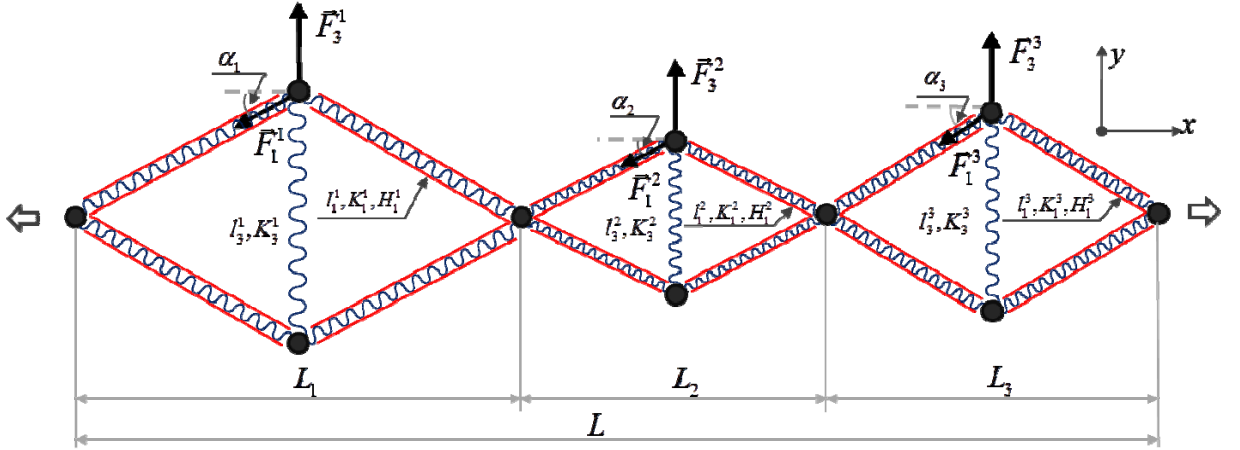

Fig. 3. Consecutive connection of three rhombic models with inclined viscous elements (triplet).

System of the ordinary nonlinear differential equations concerning variables $l_1^1(t)$, $l_1^2(t)$, $l_1^3(t)$, $l_3^1(t)$, $l_3^2(t)$, $l_3^3(t)$, $L_1(t)$, $L_2(t)$ and $L_3(t)$, describing statics and dynamics of model of consecutive connection three rhombic component:

$$\begin{cases} K_3^1 \cdot (l_{30}^1 - l_3^1) \cdot l_1^1 - (K_1^1 \cdot (l_1^1 - l_{10}^1) + H_1 \cdot \frac{d}{dt} l_1^1) \cdot l_3^1 = 0 \\ (L_1)^2 + (l_3^1)^2 - 4 \cdot (l_1^1)^2 = 0 \\ K_3^2 \cdot (l_{30}^2 - l_3^2) \cdot l_1^2 - (K_1^2 \cdot (l_1^2 - l_{10}^2) + H_2 \cdot \frac{d}{dt} l_1^2) \cdot l_3^2 = 0 \\ (L_2)^2 + (l_3^2)^2 - 4 \cdot (l_1^2)^2 = 0 \\ K_3^3 \cdot (l_{30}^3 - l_3^3) \cdot l_1^3 - (K_1^3 \cdot (l_1^3 - l_{10}^3) + H_3 \cdot \frac{d}{dt} l_1^3) \cdot l_3^3 = 0 \\ (L_3)^2 + (l_3^3)^2 - 4 \cdot (l_1^3)^2 = 0 \\ L_1 + L_2 + L_3 - L = 0 \\ K_3^1 \cdot (l_{30}^1 - l_3^1) \cdot \frac{L_1}{l_3^1} - K_3^2 \cdot (l_{30}^2 - l_3^2) \cdot \frac{L_2}{l_3^2} = 0 \\ K_3^2 \cdot (l_{30}^2 - l_3^2) \cdot \frac{L_2}{l_3^2} - K_3^3 \cdot (l_{30}^3 - l_3^3) \cdot \frac{L_3}{l_3^3} = 0 \end{cases} \quad (1.18)$$

*4 . Consecutive connection of n of rhombic blocks with inclined viscous elements.*

For the description of consecutive connection of rhombic models with known parameters of rigidity, viscosity, initial lengths of elements and full length of model, it is necessary to know values of the following variables: $l_1^1(t), l_1^2(t) \ldots l_1^n(t), l_3^1(t), l_3^2(t) \ldots l_3^n(t), L_1(t), L_2(t) \ldots L_n(t)$. Number of variables is $3n$, so the system of the equations describing consecutive connection of $n$ of rhombic models with inclined viscous elements, has to contain not less the $3n$ equations. The equations of system can be collected by the following groups: the equations, describing each rhomb separately (the equation of balance of forces in knot and the equation of geometry of a rhomb), the equation for the full length $L$ of model through the sum of lengths of each of rhombs, $L_1, L_2 \ldots L_n$ and the $n-1$ equations describing interaction of rhombs.

System of the ordinary nonlinear differential equations concerning variables, $l_1^1(t), l_1^2(t) \ldots l_1^n(t), l_3^1(t), l_3^2(t) \ldots l_3^n(t), L_1(t), L_2(t) \ldots L_n(t)$, describing dynamics for model of consecutive connection of $n$ rhombic blocks:

$$\begin{cases} K_3^1 \cdot (l_{30}^1 - l_3^1) \cdot l_1^1 - (K_1^1 \cdot (l_1^1 - l_{10}^1) + H_1 \cdot \frac{d}{dt} l_1^1) \cdot l_3^1 = 0 \\ (L_1)^2 + (l_3^1)^2 - 4 \cdot (l_1^1)^2 = 0 \\ K_3^2 \cdot (l_{30}^2 - l_3^2) \cdot l_1^2 - (K_1^2 \cdot (l_1^2 - l_{10}^2) + H_2 \cdot \frac{d}{dt} l_1^2) \cdot l_3^2 = 0 \\ (L_2)^2 + (l_3^2)^2 - 4 \cdot (l_1^2)^2 = 0 \\ \ldots \\ K_3^n \cdot (l_{30}^n - l_3^n) \cdot l_1^n - (K_1^n \cdot (l_1^n - l_{10}^n) + H_n \cdot \frac{d}{dt} l_1^n) \cdot l_3^n = 0 \\ (L_n)^2 + (l_3^n)^2 - 4 \cdot (l_1^n)^2 = 0 \\ L_1 + L_2 + \cdots + L_n - L = 0 \\ K_3^1 \cdot (l_{30}^1 - l_3^1) \cdot \frac{L_1}{l_3^1} - K_3^2 \cdot (l_{30}^2 - l_3^2) \cdot \frac{L_2}{l_3^2} = 0 \\ K_3^2 \cdot (l_{30}^2 - l_3^2) \cdot \frac{L_2}{l_3^2} - K_3^3 \cdot (l_{30}^3 - l_3^3) \cdot \frac{L_3}{l_3^3} = 0 \\ \ldots \\ K_3^{n-1} \cdot (l_{30}^{n-1} - l_3^{n-1}) \cdot \frac{L_{n-1}}{l_3^{n-1}} - K_3^n \cdot (l_{30}^n - l_3^n) \cdot \frac{L_n}{l_3^n} = 0 \end{cases} \quad (1.19)$$

## 2 . Calculation procedure

### 2.1 . Stationary case

Let the model consisting of consecutive connection of rhombs with inclined viscous elements, very slowly be exposed to influence: gradually with a negligible speed stretches to a certain final length of model $L$. In this case the viscous force created by a damper, is negligible. The system of the equations, allowing finding lengths of elements of model while the full length of connection will reach the size $L$, can be obtained from system with use of approximation $H_1 = 0, H_2 = 0 \ldots H_n = 0, \ldots$ and is equivalent to the system of the equations describing consecutive connection of rhombic models, consisting only from elastic elements. Zeroing of viscous forces will transform system of the differential equations to system of the algebraic:

$$\begin{cases} K_3^1 \cdot (l_{30}^1 - l_3^1) \cdot l_1^1 - K_1^1 \cdot (l_1^1 - l_{10}^1) \cdot l_3^1 = 0 \\ (L_1)^2 + (l_3^1)^2 - 4 \cdot (l_1^1)^2 = 0 \\ K_3^2 \cdot (l_{30}^2 - l_3^2) \cdot l_1^2 - K_1^2 \cdot (l_1^2 - l_{10}^2) \cdot l_3^2 = 0 \\ (L_2)^2 + (l_3^2)^2 - 4 \cdot (l_1^2)^2 = 0 \\ \cdots \\ K_3^m \cdot (l_{30}^n - l_3^n) \cdot l_1^n - K_1^n \cdot (l_1^n - l_{10}^n) \cdot l_3^n = 0 \\ (L_n)^2 + (l_3^n)^2 - 4 \cdot (l_1^n)^2 = 0 \\ L_1 + L_2 + \cdots L_n - L = 0 \\ K_3^1 \cdot (l_{30}^1 - l_3^1) \cdot \dfrac{L_1}{l_1^1} - K_3^2 \cdot (l_{30}^2 - l_3^2) \cdot \dfrac{L_2}{l_1^2} = 0 \\ K_3^2 \cdot (l_{30}^2 - l_3^2) \cdot \dfrac{L_2}{l_1^2} - K_3^3 \cdot (l_{30}^3 - l_3^3) \cdot \dfrac{L_3}{l_1^3} = 0 \\ \cdots \\ K_3^{n-1} \cdot (l_{30}^{n-1} - l_3^{n-1}) \cdot \dfrac{L_{n-1}}{l_1^{n-1}} - K_3^n \cdot (l_{30}^n - l_3^n) \cdot \dfrac{L_n}{l_1^n} = 0 \end{cases} \quad (2.1)$$

For obtaining of the "force - deformation" dependence it is necessary for each $L$ value from a certain set to find the corresponding value of force developed by model:

$$F = 2 \cdot F_1^1 \cdot \cos\alpha_1 = K_1^1 \cdot (l_1^1 - l_{10}^1) \cdot \dfrac{L_1}{l_1^1} \quad (2.2)$$

In the formula given above force is expressed through elements of the first model of connection, however, owing to the last equations of system describing interaction of rhombs, force can also be found from the equivalent equation expressed through elements of any connection of model:

$$F = 2 \cdot F_1^m \cdot \cos\alpha_m = K_1^m \cdot (l_1^m - l_{10}^m) \cdot \dfrac{L_m}{l_1^m} \quad (2.3)$$

Approximation of the received points $(F, L)$ will give required dependence "force - deformation".

*2.2 . Dynamic case*

Let the model consist of consecutive connection of *n* rhombs with inclined viscous elements, and then the system describing change of lengths of elements of model in time contains 3*n* equations. The analytical solution of system of the nonlinear differential equations for arbitrary number doesn't exist, for definition of dynamics of variables we have to use numerical calculations.

Dynamic characteristics are studied at sharp deformation of system (full length jump of consecutive connection of models reaches some *L* value which keeps subsequently constant). It is necessary to consider structural features of considered system. Inclined elements represent Kelvin-Voigt blocks consisting of in parallel connected spring and damper. At sharp jump of deformation the damper is frozen (approximation of infinite viscosity) therefore at the time of breakthrough the length of inclined elements remain invariable. It is a basic point to start iterative process of the computing scheme.

So, at the time of sharp deformation of system length the inclined elements keep initial values $l_1^1 = l_{10}^1$, $l_1^2 = l_{10}^2$ … $l_1^n = l_{10}^n$. From a geometrical relation for each rhomb in connection we will express length of a cross element through length of the inclined. For *m*-th rhomb the relation becomes:

$$l_3^m = \sqrt{4 \cdot (l_1^m)^2 - (L_m)^2} \quad (2.4)$$

Further we will unite the equations describing communications of rhombs with each other, and the equation representing equality of full length of connection to the sum of lengths of models of rhombs. Writing down expression for cross length $l_3^m$ through length of an inclined element $l_1^m$ and model length $L_m$, we will receive system of the equations:

$$\begin{cases} L_1 + L_2 + \cdots L_n - L = 0 \\ K_3^1 \cdot \dfrac{(l_{20}^1 - \sqrt{4\cdot(l_2^1)^2 - (L_1)^2})\cdot L_1}{\sqrt{4\cdot(l_2^1)^2 - (L_1)^2}} - K_3^2 \cdot \dfrac{(l_{20}^2 - \sqrt{4\cdot(l_2^2)^2 - (L_2)^2})\cdot L_2}{\sqrt{4\cdot(l_2^2)^2 - (L_2)^2}} = 0 \\ K_3^2 \cdot \dfrac{(l_{20}^2 - \sqrt{4\cdot(l_2^2)^2 - (L_2)^2})\cdot L_2}{\sqrt{4\cdot(l_2^2)^2 - (L_2)^2}} - K_3^3 \cdot \dfrac{(l_{20}^3 - \sqrt{4\cdot(l_2^3)^2 - (L_3)^2})\cdot L_3}{\sqrt{4\cdot(l_2^3)^2 - (L_3)^2}} = 0 \\ \cdots \\ K_3^{n-1} \cdot \dfrac{(l_{20}^{n-1} - \sqrt{4\cdot(l_2^{n-1})^2 - (L_{n-1})^2})\cdot L_{n-1}}{\sqrt{4\cdot(l_2^{n-1})^2 - (L_{n-1})^2}} - K_3^n \cdot \dfrac{(l_{20}^n - \sqrt{4\cdot(l_2^n)^2 - (L_n)^2})\cdot L_n}{\sqrt{4\cdot(l_2^n)^2 - (L_n)^2}} = 0 \end{cases} \quad (2.5)$$

The received set is system of the nonlinear algebraic equations for full lengths of rhombs, $L_1, L_2 \ldots L_n$..

Having solved the system of equations and having found $L_1, L_2 \ldots L_n$, it is possible to find numerical value of lengths of cross elements, $l_3^1, l_3^2 \ldots l_3^n$. On the following step we count values of length derivatives of inclined elements, $\frac{d}{dt}l_1^1, \frac{d}{dt}l_1^2 \ldots \frac{d}{dt}l_1^n$, expressions for which turn out from the equations connecting inclined and cross force of a certain rhomb. For *m*-th rhomb expression for a derivative becomes:

$$\frac{d}{dt}l_1^m = \frac{1}{\mu_m} \cdot \left[\frac{\mu_3^m \cdot (l_{30}^m - l_3^m) \cdot l_1^m}{l_3^m} - K_1^m \cdot (l_{10}^m - l_1^m)\right] \quad (2.6)$$

The final stage is a choice of new approach for transition to the following step of iteration. For finding of new values of inclined lengths we use Euler's method. For any time-point *t* and for *m*-th rhomb of consecutive connection new approximation turns out by addition of the product of a step of iteration $dt$ on value of a derivative to the current value:

$$l_1^m(t + dt) = l_1^m(t) + dt \cdot \frac{d}{dt}l_1^m \quad (2.7)$$

Having received a set of values, $l_1^1, l_1^2 \ldots l_1^n$, we carry out transition on the following step of iteration and algorithm becomes closed: consistently the system of the equations for $L_1, L_2 \ldots L_n$ is solved, then there are found the lengths of cross elements, $l_3^1, l_3^2 \ldots l_3^n$, derivatives $\frac{d}{dt}l_1^1, \frac{d}{dt}l_1^2 \ldots \frac{d}{dt}l_1^n$ are calculated, and on the basis of the received values on this step new approximation pays off, $l_1^1, l_1^2 \ldots l_1^n$.

At the solving of system of the algebraic nonlinear equations the specified computing scheme is universal and allows calculate dynamics of consecutive connection of any number of rhombs.

## 3 . Results.

Being guided by the above reasons, at the association of rhombic models in a doublet we won't lose sight of behavior of each model separately and we will try to allocate that essentially new that the component isn't reduced to set of separate characteristics.

### 3.1 . Nonmonotonic relaxation of deformation of a non-uniform doublet

Investigating dynamics of system, in a time-point $t = 0$ we will instantly stretch a doublet for $\Delta L$%. If length of consecutive connection in rest is $L_0$, after sharp stretching length of a doublet becomes equal $L = (1 + \frac{\Delta L}{100}) \cdot L_0$. Representing result of numerical calculation of dependence on time of a certain element of a doublet parameters in a graphic look (dynamic curve), we will provide in passing the schedule of dynamics of the same element in isolation, including identical deformation of the models.

Before starting direct numerical experiment, we will make mental experiment. Let's consistently connect two identical rhombs of equal elements length, stiffness and viscosity. It will appear that the equation of communication of two models with equal corresponding parameters degenerates into identity. The system of the equations breaks up to closed subsystems for each rhomb. It means that each rhomb in connection behaves precisely as in isolation. Identical rhombs "don't feel" influence of each other. This conclusion is fair for any number of consistently connected rhombs. The uniform system is completely determined by its only component.

Let's consider non-uniform system. Heterogeneity can be geometrical. That means different components have different lengths; it can be entered by distinction of parameters of rigidity; at last, viscosity of rhombs can accept various values. We will be limited to heterogeneity introduction on one of parameters, for example, rigidity of inclined elements:

Let's stretch a doublet for 24% :

The system of the equations describing dynamics of the connection contains as variables the length of elements. In the analysis of behavior of system from absolute sizes (lengths) we will pass to relative ones (the deformations). So, deformation of an inclined element:

$$\varepsilon_1^t = \frac{l_1^t - l_{10}^t}{l_{10}^t}, \tag{3.1}$$

deformation of a cross element:

$$\varepsilon_3^t = \frac{l_{30}^t - l_3^t}{l_{30}^t} \tag{3.2}$$

Length of an inclined element any time-point isn't less than initial length, and length of a cross element, on the contrary, in any time-point is no more than initial length. Therefore in order to operate with positive values at the description of deformation, in a formula for deformation of an inclined element from the current value it is subtracted initial, and for deformation of a cross element it is vice versa.

In fig. 4 the deformation of an inclined element of the first rhomb of the doublet with greater value in rigidity differs significantly from the second, together with the schedule of deformation of a rhomb in isolation with the same parameters. Qualitatively different behavior of a rhomb in a doublet unlike an isolation case is represented. Deformation of an inclined element of the isolated model decreases not linearly (close to an exponent), but monotonously eventually strives for the stationary value. Unlike it, in a doublet the relaxation of an element is nonmonotonic so deformation comes to static value, at first increasing and overcoming a certain maximum (Fig. 4).

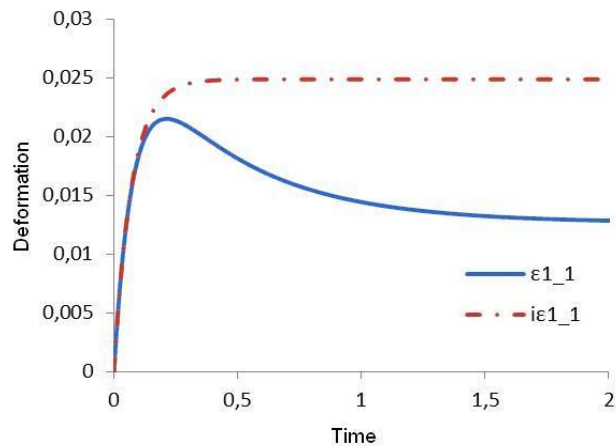

Fig. 4. Dynamic curve of inclined element deformations of a rhomb in a doublet (the continuous line) and a rhomb in isolation (dashed line). Curves are constructed for a rhomb with great value of rigidity

Let's discuss in more detail what exactly happens to each element of system in a dynamic mode, beginning with a case of model of a rhomb in isolation. In case of instant stretching the inclined element doesn't change the length (Kelvin-Voigt block is "frozen") while the cross spring is sharply reduced. The full length of system coinciding in this case with full length of a rhomb, remains invariable. Eventually the transition happens: inclined elements gradually stretch at the expense of the elastic force of the cross spring, seeking to return it to an initial condition. Over time intensity of transients falls down, and the system comes to the steady state.

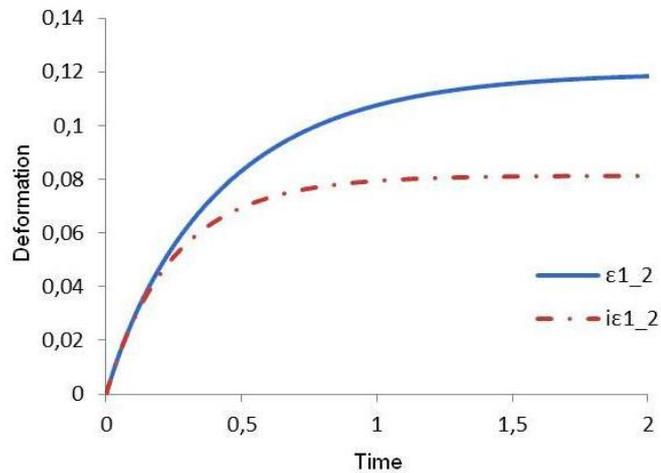

Fig. 5. Dynamic curve of inclined element deformations of a rhomb in a doublet (the continuous line) and a rhomb in isolation (dashed line). Curves are constructed for a rhomb with smaller value of rigidity

As inclined and cross lengths are connected with the full length of model by geometrical relation, knowing how change any two of them, we can define that happens to the third.

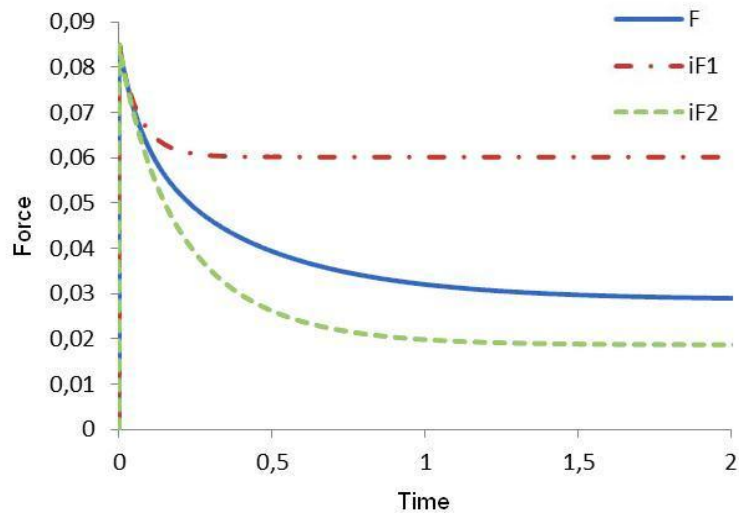

Fig. 6. The dynamic curves of forces developed by a doublet (a continuous curve) and each of rhombs in isolation.

For a rhomb with great value of rigidity deformation of an inclined element comes to smaller stationary value, than in isolation, and the full length of a rhomb during a relaxation decreases, so stationary value for deformation of a cross element of a rhomb in a doublet will strive also for smaller value, than in isolation. For softer rhomb in a doublet the opposite situation will be fair.

The curve of force developed by a doublet doesn't coincide with one of the curves of force for the isolated rhombs and can't be obtained by calculation of an arithmetic average (fig. 6).

*3.3. Criterion of existence of a nonmonotonic relaxation of deformation*

The phenomenon of a nonmonotonic relaxation of deformation can be observed in the compound system containing not less than two components. Besides, components of system have to differ though on one of parameters, modeling heterogeneity. The considered phenomenon takes place at any, even as much as small, distinction in parameters. It is possible to consider as criterion of existence of a nonmonotonic relaxation effect the presence of negative values of a derivative $\frac{dl_4}{dt}$ (or that is equivalent, $\frac{d\varepsilon_4}{dt}$) any of inclined elements as a part of a consecutive doublet. Negative values of a derivative indicate the current reduction of an inclined element in certain time-points provided that stretching (fig. 7) is typical.

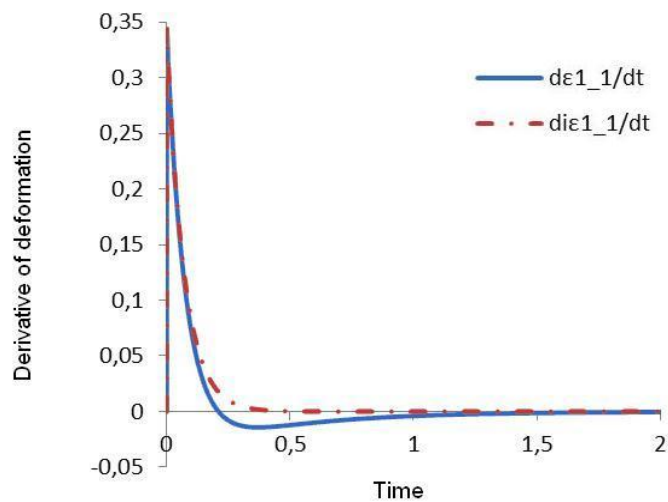

Fig. 7. Change in time of deformation derivative of an inclined element of the rhomb showing a nonmonotonic relaxation (the continuous line), in a doublet and rhomb in isolation (dashed line)

On system phase portrait condition of existence of nonmonotonic relaxation effect is a "back turn» of phase curve (fig. 8).

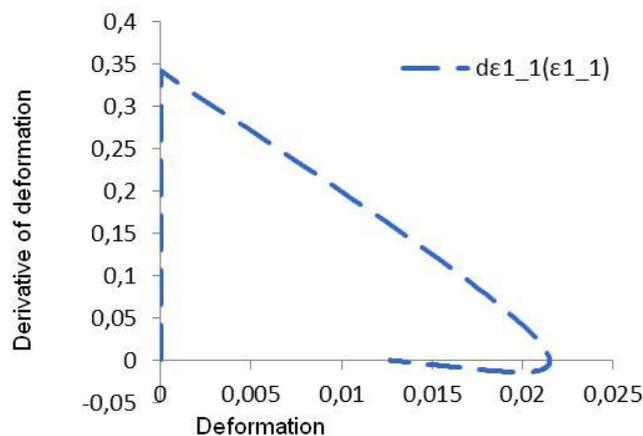

Fig. 8. Phase curve of an inclined element of a rhomb of the doublet showing a nonmonotonic relaxation

*3.4 . Step deformation*

Let's pass now from research of dynamics of model in reply to single jump of length at *t* = 0 to external deformation of the model which has been set in the form of several steps. Usually consider equal in size and equidistant on time from each other steps to analyze influence of the current level of deformation (step number) on relaxation characteristics of system, for example, time of a relaxation of tension (force) [10, 18].

Let's subject consecutive connection of two rhombic models (table 2) to step deformation: the doublet stretches on rather initial length at each step. We will be limited to consideration of the first five steps.

For a rhomb possessing great value of rigidity of an inclined spring, at each step of deformation the effect of a nonmonotonic relaxation is shown, and in process of deformation growth (with step number) the difference between value of deformation increases in a point of a maximum and stationary value (fig. 9). Also in process of growth of deformation the difference between the next stationary values grows. The last statement is fair for deformation and an inclined and cross element of each rhomb of a doublet.

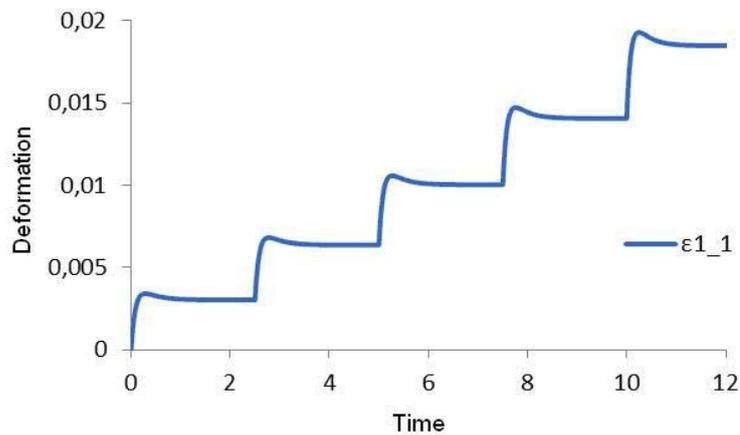

Fig. 9. Step deformation of an inclined element of a rhomb of the doublet possessing great value of rigidity

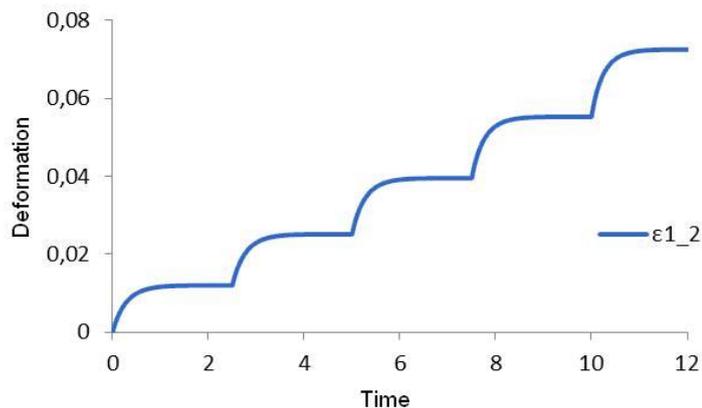

Fig. 10. Step deformation of an inclined element of a rhomb of the doublet possessing smaller value of rigidity

At step external impact on system the curves of cross deformations at small values (at initial steps) practically coincide with each other while with growth of value of deformation the difference becomes more and more obvious (fig. 11).

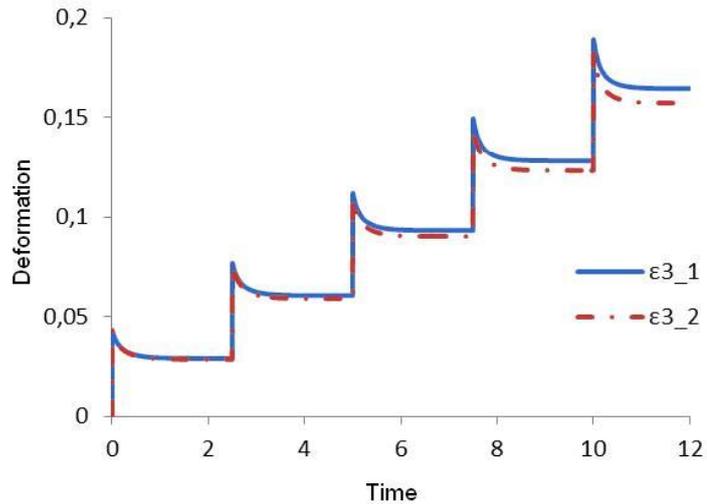

Fig. 11. Step deformations of cross elements of consecutive connection of models

It should be noted thus that the schedule of force developed by a doublet (fig. 12), corresponds to growth with increase of value of deformation of the relaxation time observed in experiment. It will be coordinated with results of research of stress relaxation in various isolated two-dimensional [10] and three-dimensional [11] models. The curve is given in the same picture for the force calculated in approach to a stationary case so that at certain time-points necessary stretching was provided. The view of a stationary curve precisely reproduces the corresponding results received in a dynamic mode, on small deformations (at initial steps), showing a small divergence with increase in deformation.

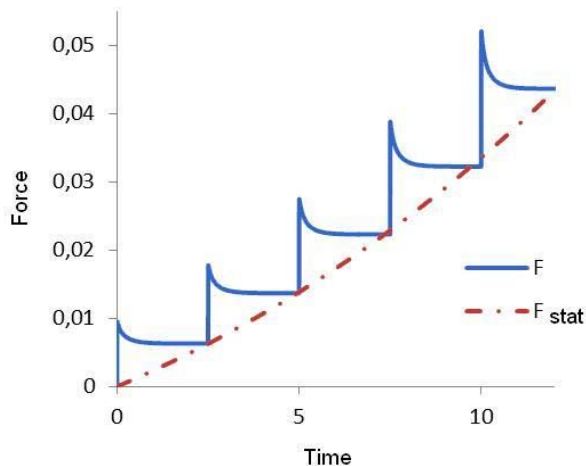

Fig. 12. The view of force developed by a doublet at step deformation (the continuous line), and the stationary curve of a doublet given to appropriate times (dashed line)

Nonlinear dependence of a relaxation of deformation on step number at step stretching is well visible on a phase portrait of the system (fig. 13). For the doublet model showing a nonmonotonic relaxation, increase of values of a derivative with deeper penetration into negative values area with deformation growth is observed. It corresponds to more intensive manifestation of effect of a relaxation of deformation at great levels of deformation that was already noted on the basis of fig. 9.

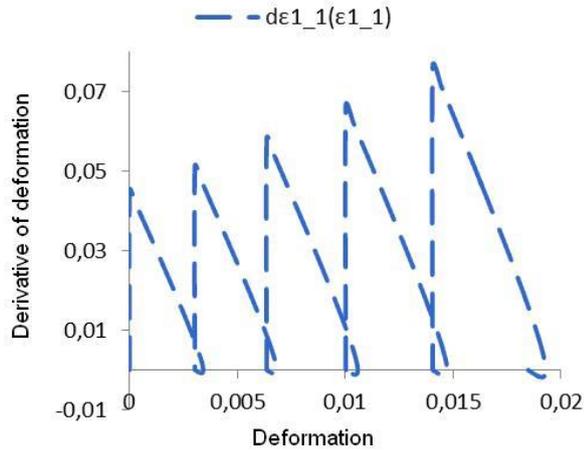

Fig. 13. Phase curve at step deformation of an inclined element of rhomb doublet showing a nonmonotonic relaxation

*3.5 . Single rectangular impulse*

Let's consider pulse external impact on full longitudinal length in model of consecutive connection of rhombs with the parameters specified in table 1. In the beginning we will stretch at a size the system from position of balance, and then, through some period, we will squeeze at the same size (to initial length). Such pulse influence serves, in particular, for testing of relaxation properties of one-dimensional models, since Maxwell's model [14].

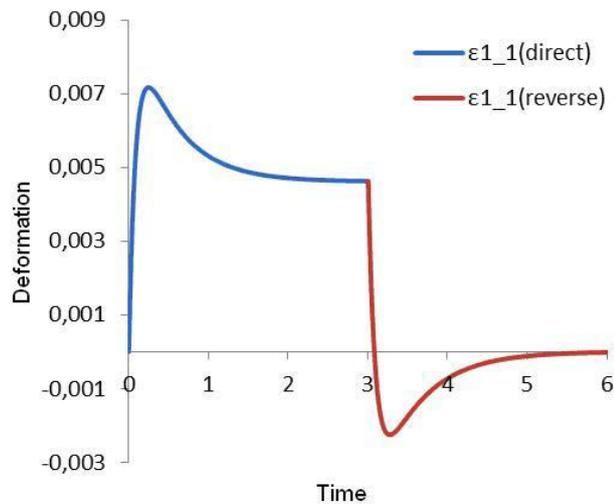

Fig. 14. Nonmonotonic relaxation of deformation of an inclined element at influence by a rectangular impulse

The nonmonotonic relaxation of deformation of an inclined element of more rigid model (fig. 14) takes place and on the forward front of an impulse (sharp stretching) and on back (sharp compression to an equilibrium state).

The relaxation of an inclined element of softer rhomb is given in fig. 15 which look is qualitatively typical for a rhomb case in isolation.

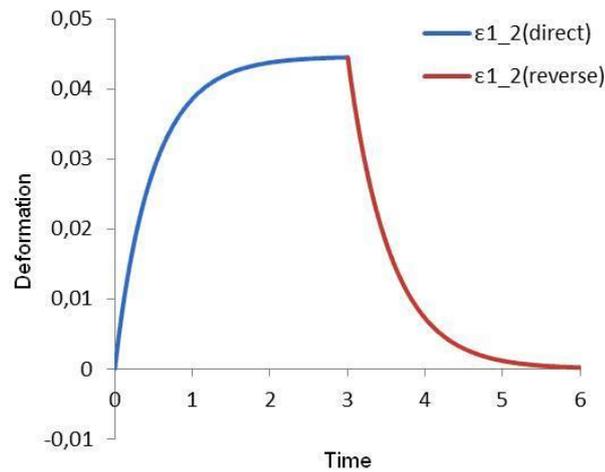

Fig. 15. Relaxation of deformation of an inclined element of softer rhomb at influence by a rectangular impulse

*3.6 . Periodic sawtooth influence*

Let's consider periodic sawtooth impact of various frequencies (conditionally 1 and 10 Hz) on the system consisting of consecutive connection of two rhombic models (parameters are specified in table 3). Amplitude of influence was chosen equal.

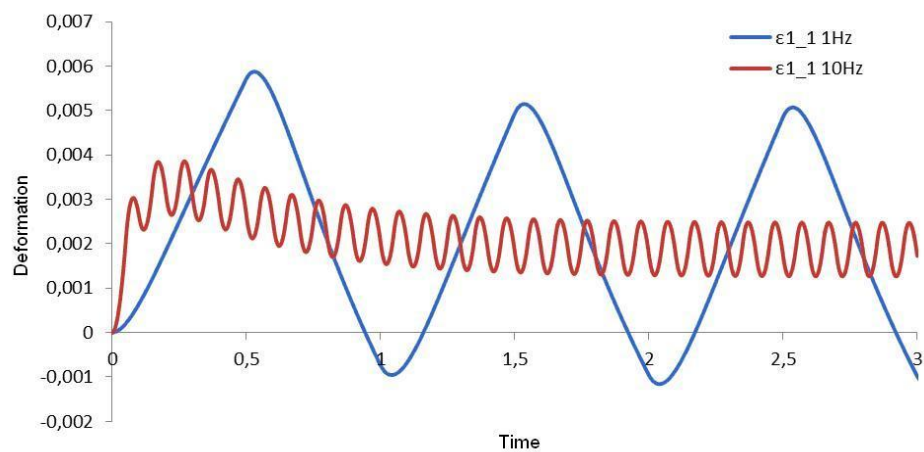

Fig. 16. Deformation of the inclined element having bigger value of rigidity, at periodic saw-tooth influence of different frequency

With a frequency of 1 Hz the system reaches stationary value after one period since the beginning of periodic influence while with a frequency of 10 Hz the system doesn't manage to adapt for changes so quickly. At rather fast external influence in case of more rigid rhomb achievement of a steady state of deformation of an inclined element happens not monotonously, that is through overcoming of a certain maximum (fig. 16). Not monotony of the average line is meant.
In case of softer rhomb deformation of an inclined element reaches stationary value monotonously (fig. 17). Monotony of achievement of stationary value will be fair as well for cross elements of a rhomb.

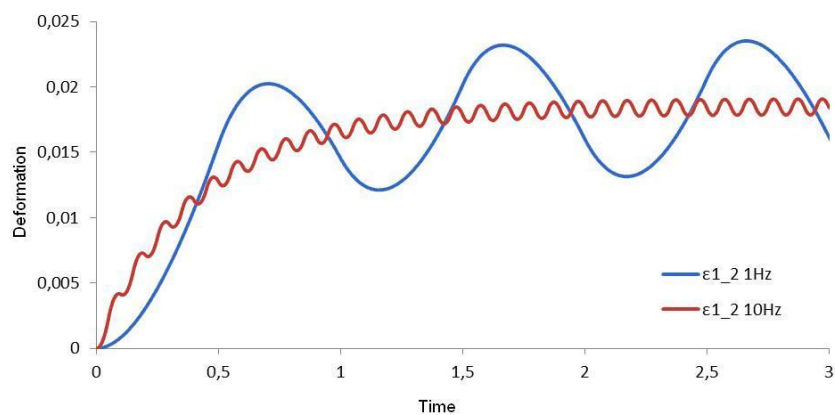

Fig. 17. Deformation of the inclined element having smaller value of rigidity, at periodic saw-tooth influence of different frequency

### 3.7. Periodic sinusoidal influence

Research of quasistationary properties of consecutive connection includes finding of a hysteresis loop of a system. Let put on sinusoidal influence with an amplitude move $\Delta L = 25\%$ on consecutive doublet (parameters are specified in table 3).

The hysteresis loop (fig. 18) has a "banana-like" shape, characteristic for the majority of biological tissues. Influence of degree of heterogeneity on a hysteresis loops was not revealed.

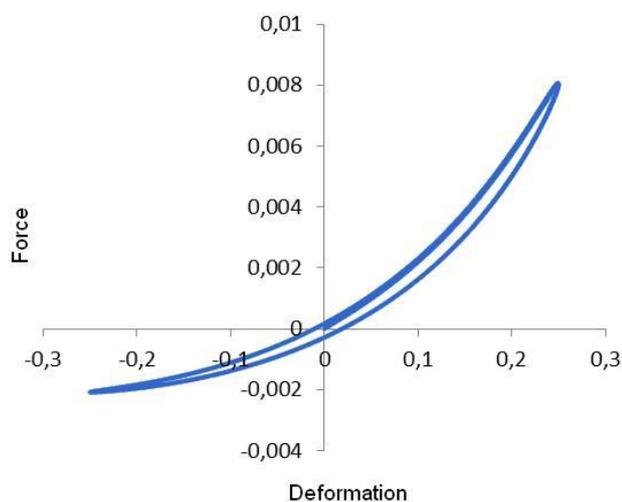

Fig. 18. Hysteresis in doublet model at periodic sinusoidal influence

### 3.8. Nonmonotonic relaxation of deformation in doublet model with cross viscous elements

The nonmonotonic relaxation of deformation takes place not only in a consecutive doublet of rhombic models with inclined viscous elements, but also in a doublet of models with cross viscous elements. In this case the rhombic model which inclined elements are elastic, and cross represents Kelvin-Voigt block (parallel connection of a spring and a damper) acts as structural unit. In this case the nonmonotonic relaxation is observed for deformation of a cross element (on that element of the doublet which value is greater provided that all other mechanical and geometrical parameters are identical).

*3.9 . Nonmonotonic relaxation of deformation of blocks in consecutive connection of three rhombs (triplet).*

Up to this point the object of research was the consecutive doublet. Not evident and original result was detection of an abnormal course of a relaxation of a structural element of the separate block length (for model with inclined viscous elements the nonmonotonic relaxation of deformation is observed on an inclined element of the block, for model with cross viscous elements – on a cross element of the block). There is a question as the system will lead at further scaling, that is at increase in number of blocks.

It was investigated in numerical experiment the non-uniform system composed of three consistently connected blocks with inclined viscous elements by means of which the first results in modeling of dynamics of system at impact of one-stage deformation are obtained. Parameters of structural elements of blocks are described in table 4, block lengthening at sharp stretching was chosen.

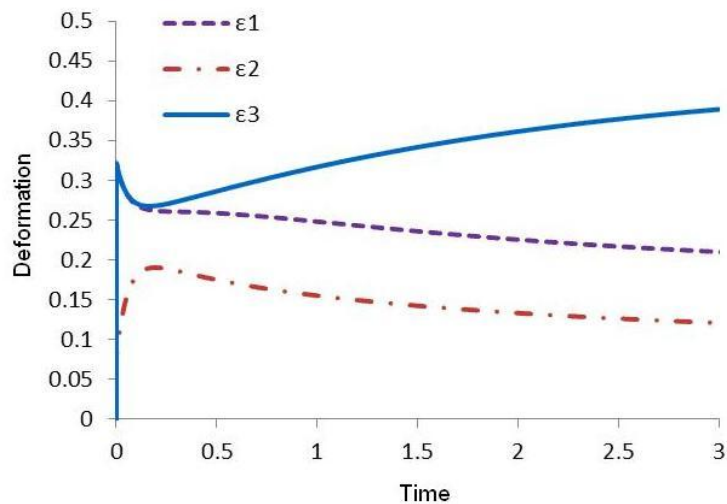

Fig. 19. Nonmonotonic relaxation of longitudinal deformation of the block in model of consecutive connection of three rhombs (triplet).

The nonmonotonic relaxation of deformation of an inclined element takes place and is observed on two blocks from three, but in the model of three rhombs there is a qualitative difference: in this case in a nonmonotonic way changes also longitudinal deformation of the block.

**Discussion**

Let's note that when approaching to research of viscoelastic properties of the structured environments and, in particular, biological soft tissues, we at the first stage were limited to consideration of Basic Elements of the tissue representing minimum by the sizes structural unit, keeping its main properties. Thus we were based on one of fundamental approaches to consideration of any question - from simple to difficult. Starting the solution of a problem, splitting of object of studying into parts and allocation of certain structural unit (analysis) appears a productive method. Various properties of part in itself, isolated from complete system are studied. Before starting describing characteristics of compound system, it is necessary to prepare base of how separate components of this system behave. Further it will allow to describe and list not simply a set of characteristics of system, and to specify emergent properties, qualitatively new that takes place only in complete system and isn't observed at a component separately. The following step is association of the shattered parts (synthesis) and their research in interaction with each other.

Uniformity of properties of system in space, i.e. an invariance of its characteristic parameters from point to point, is one of initial essential simplifications of the theory which, however, allows to move ahead considerably in its studying at the expense of many mathematical and physical theorems of conservation (of an impulse, energy and so forth) . In the practice we observe always to some extent non-uniform environments and conditions. The elementary reception for the accounting of heterogeneity is connection into a chain of elementary structural units with various geometrical, elastic and viscous properties. Studying of properties of such dimers, which we will call a parallel doublet (or a tandem), or at last triplets, will allow to find their special characteristics which will appear at further "polymerization". Heterogeneity in this case is imitated by dependence of parameters of structural units on their position in system, i.e. from number in a chain or from knot coordinates in a grid. Thus, along with nonlinearity, in one model one more property of a matter - its heterogeneity can be considered.

In case of consecutive connection of several rhombs each of them has an additional degree of freedom: length of each rhomb can change the value. Full length of system at similar influence still has to remain invariable, and therefore change of length of one block has to be compensated by change of lengths of others. In the tandem of the rhombs considered by us, the following occurs with inclined elements of different stiffness: at the time of sharp stretching the viscosity of blocks of Kelvin-Voigt of both rhombs is possible to consider as infinite that doesn't allow an inclined element to react to instant influence (owing to effect of "frozen down" of the block). Therefore total stretching at the time of breakthrough is completely carried out at the expense of compression of cross springs. In case of geometrically identical rhombs, at the time of sharp stretching the full length of both rhombs is identical and equal to that value which each of rhombs would be in isolation. However, "inclusion" of inclined elements leads to a shift of a point of connection of models. Full length of a rhomb with bigger rigidity decreases and for compensation the full length of other rhomb with smaller rigidity needs to increase. In some time-point determined by maximum of deformation curve, the elastic force of an inclined element of more rigid rhomb starts dominating, seeking to return it to an equilibrium condition, and Kelvin-Voigt block starts contracting. As a result stationary value of deformation of an inclined element of more rigid rhomb of a doublet is smaller in comparison with the corresponding value of the same rhomb in isolation. It is caused by interrelation of rhombs in the system. To support invariable the total longitudinal length of system, an inclined element of softer rhomb in a doublet, stretching more and more, comes to bigger stationary value of deformation, than in isolation.

The phenomenon, in a doublet observed only on a structural element of the block, in triplet is shown at macroscopic level – at the level of the block's longitudinal length. It is important for the accessibility of experimental check of the phenomenon of a nonmonotonic relaxation. Such qualitative difference in relaxation behavior of model with number of blocks more than two, apparently is possible to explain by the fact that in a doublet owing to constant full longitudinal length condition there are lack, or not enough, longitudinal degrees of freedom.

Thus, in the present work introductive approaches to modeling of a flat layer of the structured material were investigated. The rhombic model acted as structural unit on the basis of which the non-uniform system was represented by consecutive connection of structural elements. The response of system to various external influences was studied. So:

• The algorithm of the solution of a problem of the accounting for heterogeneity of the structured material in model of arbitrary consecutive connection of elements is developed.

• For the first time by the method of mathematical modeling conditions of emergence and features of a non-monotonic relaxation of deformation in non-uniform materials are analyzed.

• Influence of a various type of external step and periodic stretching on a nonmonotonic deformation relaxation is investigated.

Further interest presents studying of the effects arising at scaling of the system. Open question there is a problem of connection of structural elements in the transversal direction.

**Acknowledgements:** The authors are grateful to all participants of the seminars of biological motility laboratory of the Institute of immunology and physiology UrD RAS and laboratory of kinetic phenomena of the Institute for physics of metals UrD RAS for active interest to the work and valuable remarks.

## Appendix A

For more realistic accounting for the conditions of mechanical contact and edge effects it is necessary to enter the rigid coupling bar to which the model elements fasten at the edges. Introduction in coupling system the bar of $h$ height leads to modification of the geometrical equation and another record of a sine of the angle between an inclined element and a horizontal axis.

$$\sin \alpha = \frac{l_3}{2 \cdot l_1} \rightarrow \sin \alpha = \frac{l_3 - h}{2 \cdot l_1} \tag{A1}$$

$$L^2 + l_3^2 = 4 \cdot l_1^2 \rightarrow L^2 + (l_3 - h)^2 = 4 \cdot l_1^2 \tag{A2}$$

In the case of consecutive connection of two rhombs with inclined viscous elements the system of equations describing dynamics of a tandem, at introduction of the coupling bar will become:

$$\begin{cases} K_3^1 \cdot (l_{30}^1 - l_3^1) \cdot l_1^1 - \left(K_1^1 \cdot (l_1^1 - l_{10}^1) + H_1 \cdot \frac{d}{dt}l_1^1\right) \cdot (l_3^1 - h) = 0 \\ (L_1)^2 + (l_3^1 - h)^2 - 4 \cdot (l_1^1)^2 = 0 \\ K_3^2 \cdot (l_{30}^2 - l_3^2) \cdot l_1^2 - \left(K_1^2 \cdot (l_1^2 - l_{10}^2) + H_2 \cdot \frac{d}{dt}l_1^2\right) \cdot (l_3^2 - h) = 0 \\ (L_2)^2 + (l_3^2 - h)^2 - 4 \cdot (l_1^2)^2 = 0 \\ L_1 + L_2 - L = 0 \\ K_3^1 \cdot (l_{30}^1 - l_3^1) \cdot \frac{L_1}{l_3^1 - h} - K_3^2 \cdot (l_{30}^2 - l_3^2) \cdot \frac{L_2}{l_3^2 - h} = 0 \end{cases} \tag{A3}$$

## Appendix B

Table 1
Parameters of consecutive connection of two rhombs (doublet)

|  | *First rhomb $i = 1$* | *Second rhomb $i = 2$* |
|---|---|---|
| $L_i$ | 2 | 2 |
| $l_3^i$ | 2.2 | 2.2 |
| $l_1^i$ | 1.49 | 1.49 |
| $K_3^i$ | 0.1 | 0.1 |
| $K_1^i$ | 1 | 0.1 |
| $H_i$ | 0.1 | 0.1 |

Table 2
Parameters of doublet at step deformation

|  | *First rhomb $i = 1$* | *Second rhomb $i = 2$* |
|---|---|---|
| $L_i$ | 2 | 2 |
| $l_2^i$ | 2.2 | 2.2 |
| $l_1^i$ | 1.49 | 1.49 |
| $K_2^i$ | 0.1 | 0.1 |
| $K_1^i$ | 1 | 0.25 |
| $H_i$ | 0.1 | 0.1 |

**Table 3**
Parameters of doublet at sinusoidal action

|  | First rhomb $i = 1$ | Second rhomb $i = 2$ |
|---|---|---|
| $L_i$ | 2 | 2 |
| $l_2^i$ | 2.2 | 2.2 |
| $l_1^i$ | 1.49 | 1.49 |
| $K_2^i$ | 0.01 | 0.01 |
| $K_1^i$ | 1 | 0.1 |
| $H_i$ | 0.1 | 0.1 |

**Table 4**
Parameters of triplet at sinusoidal action

|  | First rhomb $i = 1$ | Second rhomb $i = 2$ | Third rhomb $i = 3$ |
|---|---|---|---|
| $L_i$ | 2 | 2 | 2 |
| $l_2^i$ | 2.2 | 2.2 | 2.2 |
| $l_1^i$ | 1.49 | 1.49 | 1.49 |
| $K_2^i$ | 0.1 | 1 | 0.1 |
| $K_1^i$ | 1 | 0.5 | 0.1 |
| $H_i$ | 0.4 | 0.05 | 0.4 |